\documentclass{article}

\usepackage{PRIMEarxiv}

\usepackage[utf8]{inputenc} 
\usepackage[T1]{fontenc}    
\usepackage{hyperref}       
\usepackage{url}            
\usepackage{booktabs}       
\usepackage{amsfonts}       
\usepackage{nicefrac}       
\usepackage{microtype}      
\usepackage{lipsum}
\usepackage{fancyhdr}       
\usepackage{graphicx}       
\graphicspath{{media/}}     
\usepackage{graphicx}
\usepackage[colorinlistoftodos,prependcaption]{todonotes}
\usepackage{regexpatch}
\usepackage{subcaption}
\usepackage{float}
\usepackage{mwe}
\usepackage{amsmath}
\title{Global urban activity changes from COVID-19 physical distancing restrictions
}

\author{
  Srija Chakraborty \\
  USRA \\
  Washington, DC\\
  \texttt{\{schakraborty\}@usra.edu} \\
   \And
  Eleanor C. Stokes \\
  NASA \\
  Washington, DC\\
   \And
  Olivia Alexander \\
  USRA \\
  Mountain View, CA\\
}

\begin{document}
\maketitle

\begin{abstract}
During the COVID-19 pandemic changes in human activity became widespread through official policies and organically in response to the virus's transmission, which in turn, impacted the environment and the economy.  The pandemic has been described as a natural experiment that tested how social and economic disruptions impacted different components of the global Earth System.  To move this beyond hypotheses, locally-resolved, globally-available measures of how, where, and when human activity changed are critically needed. Here we use satellite-derived nighttime lights to quantify and map daily changes in human activity that are atypical for each urban area globally for two years after the onset of the pandemic using machine learning anomaly detectors.  Metrics characterizing changes in lights from pre-COVID baseline in human settlements and quality assurance measures are reported. This dataset, TRacking Anomalous COVID-19 induced changEs in NTL (TRACE-NTL), is the first to resolve COVID-19 disruptions for all metropolitan regions globally, daily. It is suitable to support a variety of post-pandemic studies that assess how changes in human activity impact environmental systems.  
\end{abstract}


\section{Introduction}
In response to the global COVID-19 pandemic, national, state, and local governments instituted stringent social distancing and lock-down measures, including closing non-essential businesses, schools, entertainment venues, and public transportation to mitigate virus transmission. In addition to official measures, many people voluntarily stayed at home or otherwise altered where they worked or lived in response to the pandemic. Collectively these measures comprised the largest medical quarantine event in recorded history, and resulted in changes in air pollution \cite{venter2020covid, shi2021abrupt, fu2020impact}, secondary inorganic aerosols and the global radiative budget~\cite{sekiya2023worldwide}, energy consumption \cite{jiang2021impacts, gillingham2020short, li2023exploring}, the energy-water nexus~\cite{bhattacharya2023energy}, emissions \cite{le2020temporary, liu2020near,forster2020current, kumar2022impact, yuan2023progress, guevara2023towards}, nighttime atmospheric chemistry~\cite{yan2023increasing}, wetland methane emission \cite{peng2022wetland} and human-wildlife interactions \cite{manenti2020good, naidoo2021socio, cukor2021different, tucker2023behavioral}. 

Because of its wide-reaching impacts on the environment, the pandemic has been framed as a natural experiment for observing the world devoid of certain anthropogenic pressures, or an "anthropause" \cite{diffenbaugh2020covid, gorris2021shaping, bates2020covid, rutz2020covid}. Studies have explored how this natural experiment impacted environmental factors such as air quality~\cite{miyazaki2021global, he2024covid} which in turn impacted human health~\cite{he2024covid}. Satellite data has been used to track anthropause-induced changes in the environment like air quality \cite{venter2020covid, cooper2022global}, carbon emissions \cite{liu2020near, ray2022impact,peng2022wetland, zhang2023quantifying}, secondary inorganic aerosols~\cite{yan2023increasing}  and soil organic carbon \cite{ray2022quantifying}. However, in addition to collecting data on environmental impacts, global data about how and where human activity patterns changed during the pandemic are needed to advance understanding of the long term response of Earth systems.  Mobility datasets, like those from Google Maps \cite{aktay2020google} and Apple maps \cite{apple2020covid}, provide some insights on changes in human travel during the pandemic, but these datasets are less informative on local scale patterns since data from most countries is aggregated at the national level or only presented for few of the largest cities.  Case studies of human activity patterns during the pandemic have indicated there were high spatial heterogeneities within countries \cite{lee2020human, stokes2022tracking,hou2021intracounty}, so national scale mobility data is of limited use for characterizing local disruptions and resulting changes in human presence.  Similarly, numerous repositories of government policies have been compiled \cite{cheng2020covid,hale2021global,desvars2020structured,piccoli2021citizenship,zheng2020hit} but government policies are often disassociated from the human activity patterns they are meant to influence \cite{kim2021impact,smyth2022fading}. 

Standardized, global, highly resolved spatio-temporal data about how human activity changed during the pandemic remains a significant gap --- one where satellite data can add insight.  Remotely-sensed nighttime lights (NTL) are routinely collected globally at the neighborhood scale on a daily basis, characterizing a mix of the social, cultural, and economic dynamics within human settlements. Nightlight dynamics during COVID-19 have captured business closures \cite{hayakawa2022effective,lan2021quantifying} and changes in traffic associated with "stay at home" orders \cite{jechow2020evidence}, as well as changes in how holidays like Ramadan were observed \cite{stokes2022tracking, wang2022human}.  Here we monitor daily NTL globally over 9031 urban settlements for the three years before and two years after the onset of the pandemic, to construct a spatio-temporal data repository, TRacking Anomalous  COVID-19 induced changEs in NTL (TRACE-NTL) recording the timing and degree of change in human activity.

Our dataset is built from satellite-derived nighttime lights (NTL) captured from the Visible Infrared Imaging Radiometer Suite Day/Night Band (VIIRS-DNB) onboard the Suomi-NPP 
polar-orbiting satellite platform. Using the VNP46A2 daily gap-fill product, we create a stable time series for each settlement devoid of natural aberrations from seasonal vegetation, terrain effects, snow, atmospheric variation, and moonlight \cite{roman2018nasa}, and aggregate NTL within each settlement. Machine learning models are then used to characterize typical patterns in nightlights \cite{chakraborty2023adaptive-ts} for each settlement pre-2020 (e.g. patterns of light change associated with urbanization or infrastructure investment, the urban seasonal holiday signal, and the prevalence of electricity instability). We then apply these machine learning models to pandemic era data to detect anomalous changes in NTL radiance. The resulting dataset includes multiple measures characterizing the impact of COVID-19 on the NTL signal--including the magnitude of the change in radiance, the direction of change, its duration, and several quality assurance indices. These describe how anomalous the change was to facilitate appropriate use of the measures by a wide swath of technical and non-technical stakeholders.  With this database, we provide a complete, consistently-measured, openly-available record of urban light patterns globally, as well as the timing of recovery to "normal" NTL levels. We expect these data to inform global economic and environmental analyses, assessing the long term impacts from the pandemic. 

The dataset is not a direct measure of human movement or economic activity and thus has some limitations for characterizing either. A large portion of the lighting in cities is comprised of streetlights, which in most places have remained lit throughout the pandemic. However, a portion of nightlights in human settlements are ephemeral, and light reductions are associated with the closure of businesses, the slowing of transportation activity and travel, the cancelling of cultural and social events, and the shuttering of industrial facilities, all of which have local and global economic and environmental impacts. Though NTL is not a perfect proxy for any single anthropogenic activity, it is one of the only measures of human presence that is collected in a standardized way with comprehensive coverage across the globe.  With tens of thousands of human settlements worldwide, all highly interconnected through industrial supply chains and mobility networks, and each with different control strategies and different time schedules for closing and opening their economies, a dataset that can capture this variety has the potential to advance understanding of the multi-faceted, systemic impact of the pandemic on human-environment interactions.

\section*{Data and Results Summary}
\begin{figure}[h]
    \centering
    \begin{subfigure}[b]{0.49\linewidth}        
        \centering
        \includegraphics[width=\linewidth]{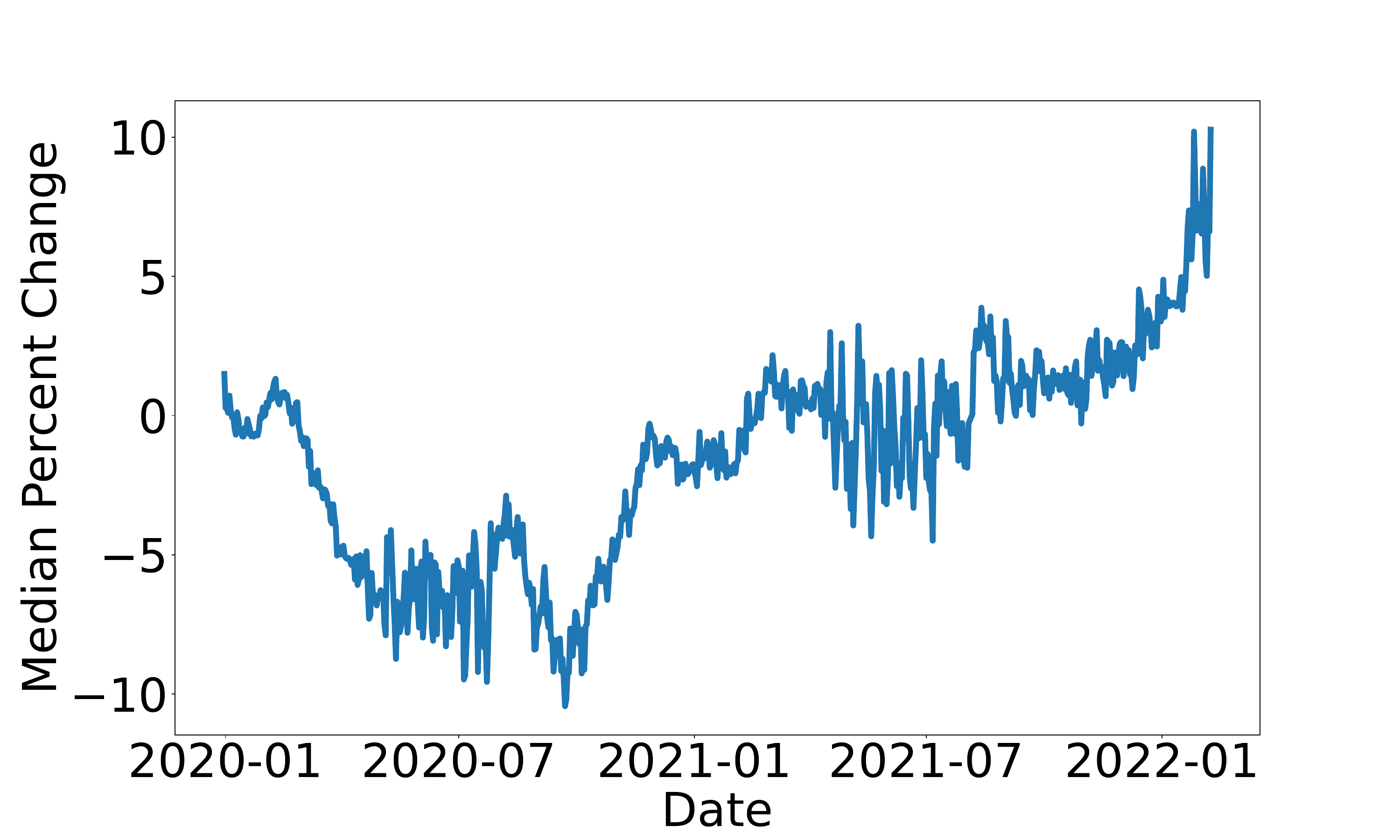}
        \caption{Global}
        \label{fig:chA}
    \end{subfigure}
    \begin{subfigure}[b]{0.49\linewidth}        
        \centering
        \includegraphics[width=\linewidth]{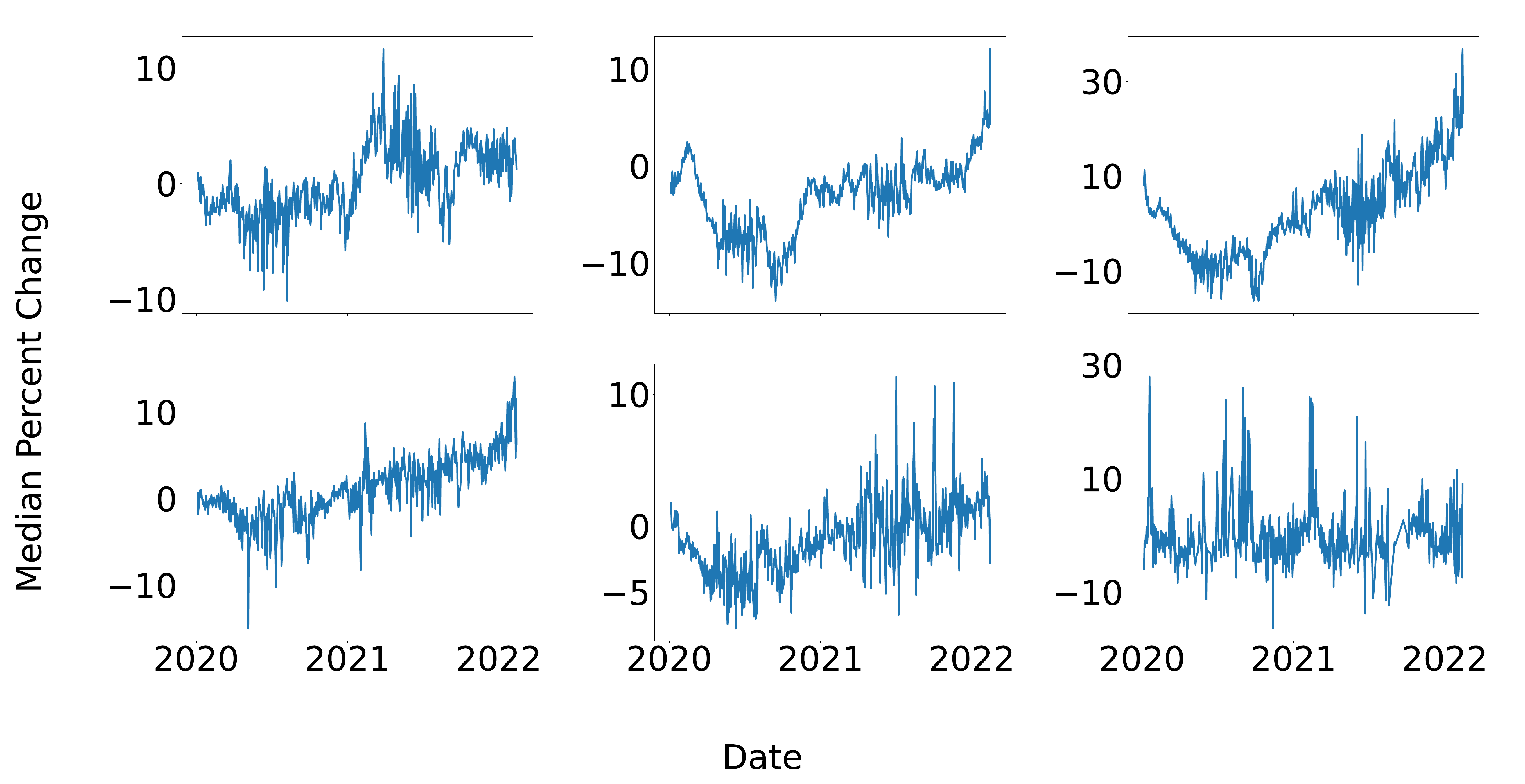}
        \caption{(top left to bottom right) Europe, Asia, Africa, North America, South America, Oceania}
        \label{fig:chB}
    \end{subfigure}
    \caption{Median percent change between observed NTL and forecasts derived from normal pre-COVID-19 years. Negative change shows decline in NTL. Figures show (a) global, (b) continental breakdown of patterns. }
    \label{fig:perchange}
\end{figure}

We highlight the key trends captured by NTL through time-series analysis over 9031 urban areas and summarize the results at multiple scales. Changes in the NTL time-series are detected through two approaches: (1) using a time-series forecast based solely on pre-2020 NTL data, to determine the deviation and note recovery from a “normal”  pre-pandemic baseline, and (2) by allowing the baseline to adjust and adapt to new NTL observations within 2020-2022, to track waves of responses to new COVID mitigation measures and outbreaks within the pandemic period. For (1) we use STL decomposition~\cite{cleveland1990stl} of pre-pandemic NTL observations to forecast NTL patterns through 2020-2022,and then measure the deviation of COVID-era observations from that forecast. For (2) we use unsupervised deep learning models~\cite{chakraborty2023adaptive-ts} to learn the natural variation in pre-COVID urban NTL data, and then continuously predict NTL observations within the 2020-2022 pandemic period based on the closest previous months of NTL data. We then identify where observations meaningfully deviate from those predictions as change points.  Importantly, with this approach, the models quickly adapt to a new baseline after an anomalous event and monitor further disruptions from it.  Both approaches learn city-specific models from pre-COVID-19 years and apply those models to monitor the given urban area, allowing the analysis to be globally scalable and adaptive to the NTL variations across cities (see Methods). 

Figure~\ref{fig:chA} shows the daily median of percentage change in NTL from the baseline forecasted from pre-pandemic levels  for all urban areas globally (detected using the decomposition approach (1)). We observe an abrupt negative median percent change (decline in NTL compared to pre-COVID years) in the forecast starting in early March 2020, which briefly subsides in the summer months, and then drops to the highest global decline level of -10\% in September 2020. An equally abrupt recovery is observed in late fall, with NTL levels eventually reaching the baseline forecasted from pre-pandemic levels in March 2021. The median percentage change hovers around zero and then slowly resumes the normal pattern of increasing from August 2021. Global NTL levels have long trended upwards, due to the combination of population increases, urbanization, and electrification gains~\cite{li2020harmonized}, and without the effects of global scale disruptions like the pandemic, NTL is expected to continue to increase.
Figure~\ref{fig:chB} shows a continental scale breakdown of the global median percent change, with declining NTL observed in early 2020 consistently across the continents. The magnitude and duration of light loss associated with the onset of the pandemic varies across continents, with Asia, Africa, and South America showing the most dramatic and sustained decreases in NTL levels over predicted normal levels.

\begin{figure}[h]
    \centering
    \begin{subfigure}[b]{0.45\linewidth}        
        \centering
        \includegraphics[width=\linewidth]{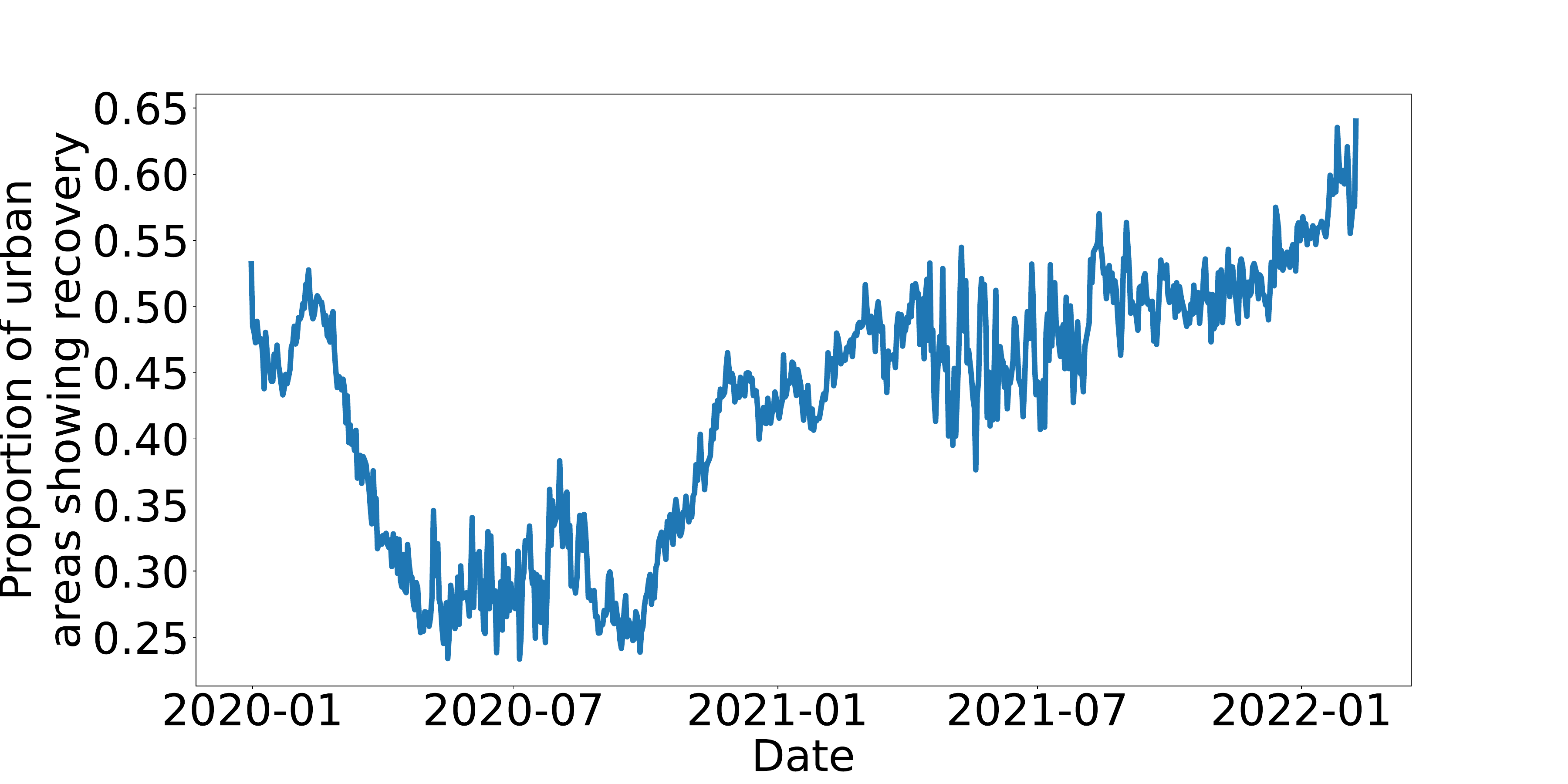}
        \caption{Ongoing global recovery}
        \label{fig:rec_g}
    \end{subfigure}
    \begin{subfigure}[b]{0.5\linewidth}        
        \centering
        \includegraphics[width=\linewidth]{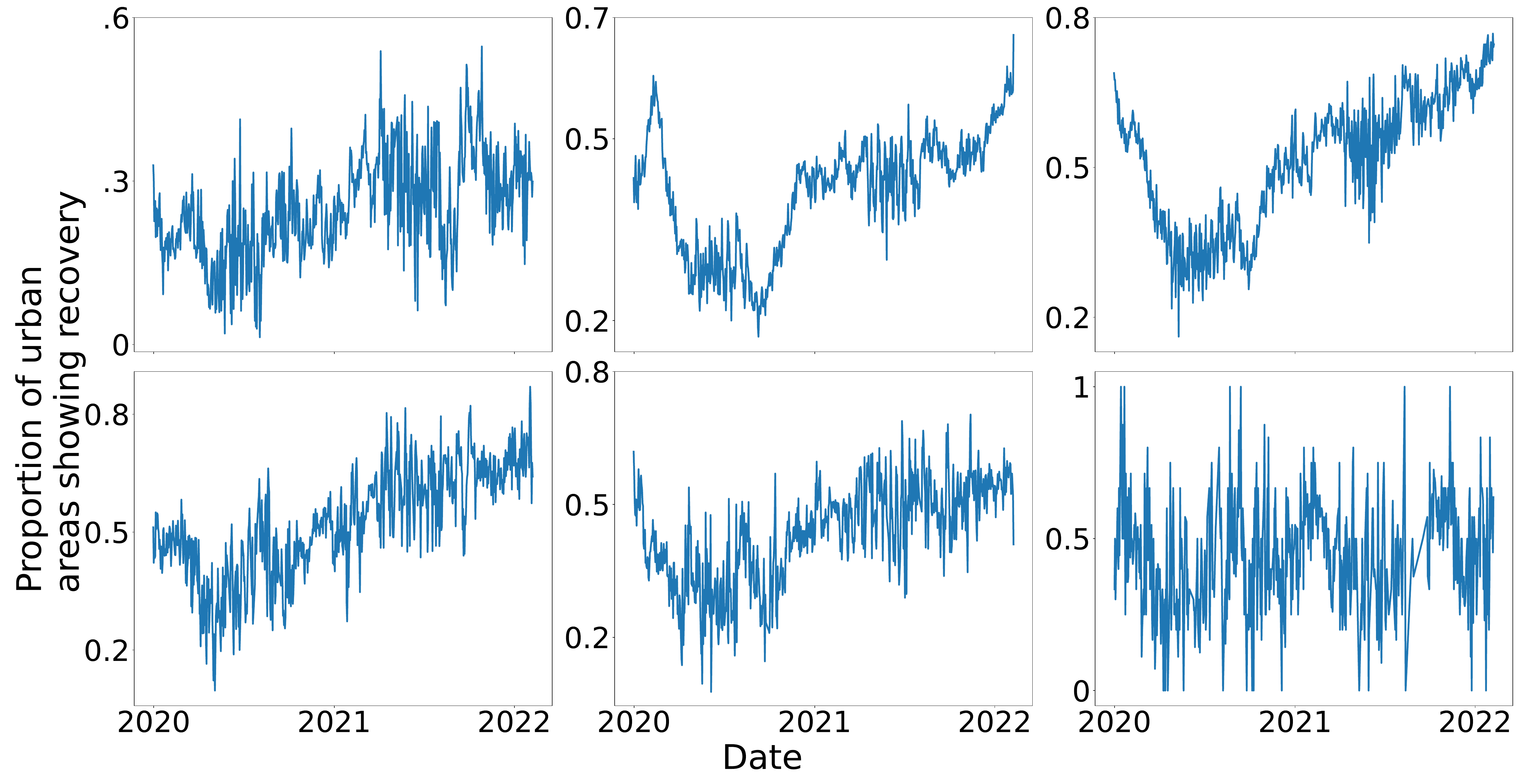}
        \caption{Ongoing recovery (top left to bottom right) Europe, Asia, Africa, North America, South America, Oceania}
        \label{fig:rec_c}
    \end{subfigure}
    \caption{Proportion of urban areas showing recovery, which is marked by observed NTL after the onset of COVID-19, exceeding the forecast derived from normal pre-COVID-19 years. Figures show (a) global, (b) continental breakdown of recovery patterns.}
    \label{fig:rec_stl}
\end{figure}

Recovery can also be identified by comparing NTL forecasts derived from pre-pandemic years with observed NTL (see Methods). Recovery is reached when the observed NTL meets or exceeds the forecasts derived from pre-pandemic years. Figure~\ref{fig:rec_stl} shows ongoing recoveries after the pandemic-induced decline in NTL, measured using the fraction of urban areas where recovery has been reached. With the onset of COVID-19, approximately 75\% of global urban areas had lower NTL level compared to pre-COVID-19 years. By August 2021 more than half of global urban areas had NTL levels indicating recovery, with that proportion increasing through 2022. 

\begin{figure}
    \centering
    \begin{subfigure}[b]{0.49\linewidth}        
        \centering
        \includegraphics[width=\linewidth]{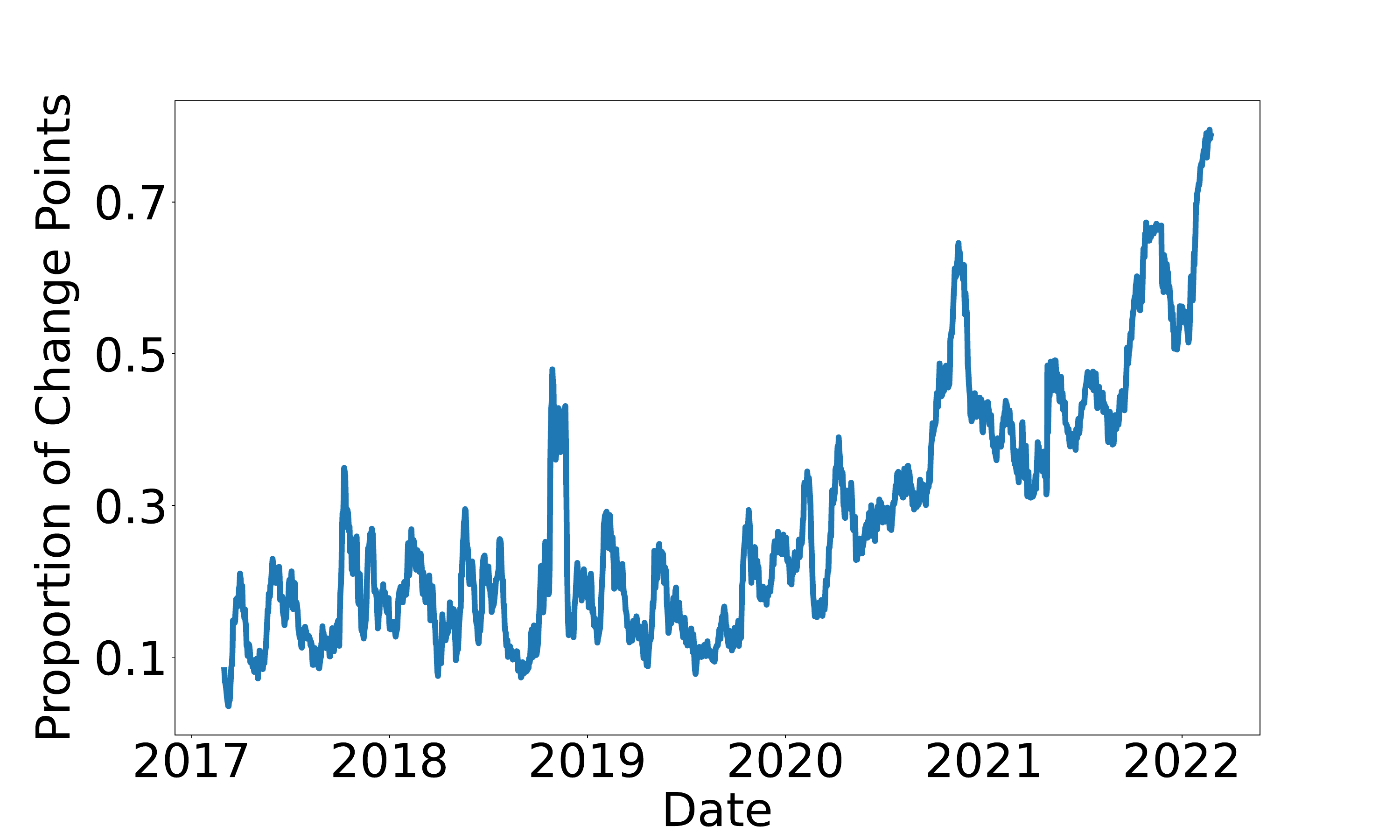}
        \caption{Global}
        \label{fig:prA}
    \end{subfigure}
    \begin{subfigure}[b]{0.49\linewidth}        
        \centering
        \includegraphics[width=\linewidth]{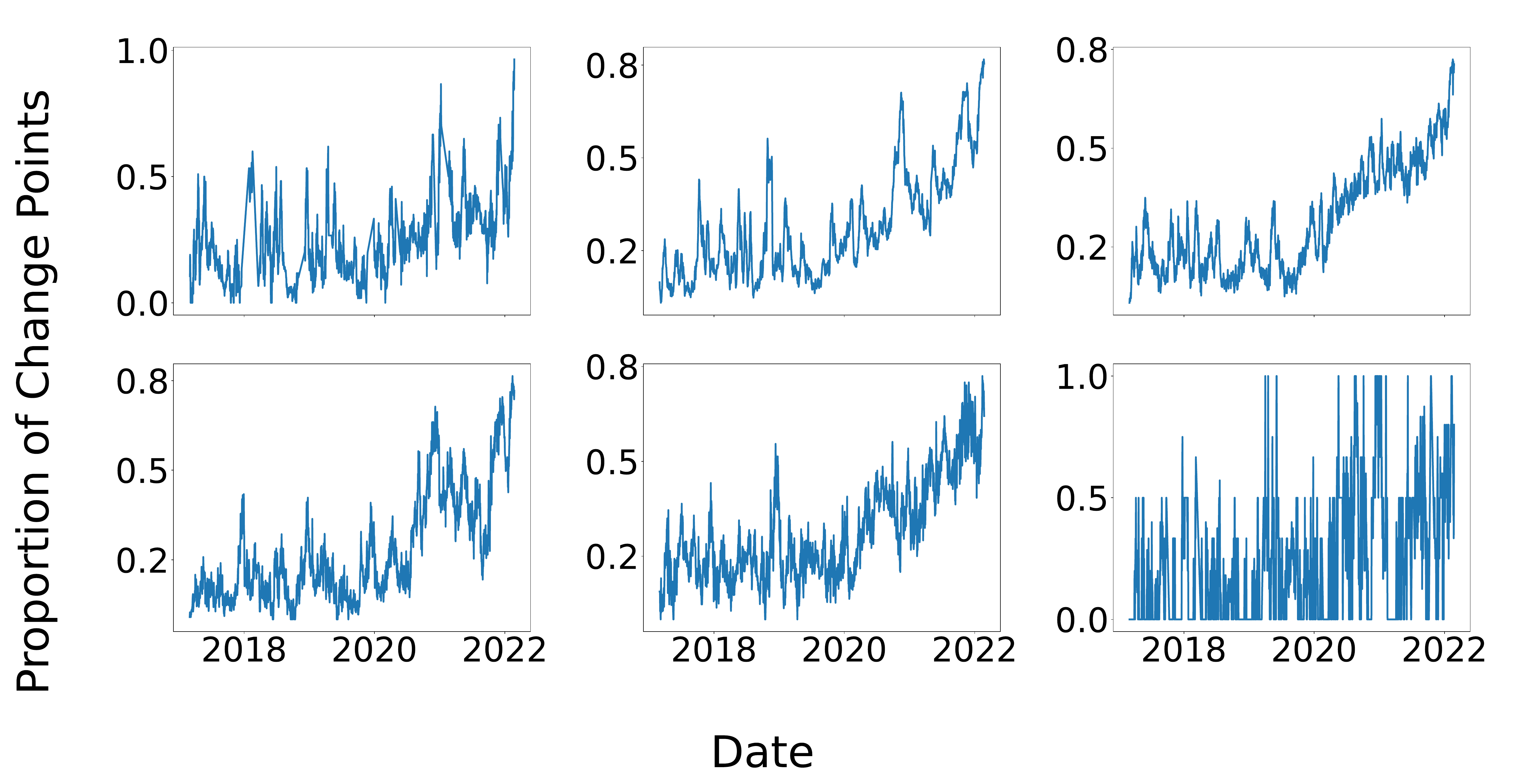}
        \caption{(top left to bottom right) Europe, Asia, Africa, North America, South America, Oceania}
        \label{fig:prB}
    \end{subfigure}
    \caption{Proportion of urban areas with anomalous change detected using windows of NTL data using machine learning on each day from Jan 2017-Feb 2022, with a continued increase from mid-March 2020 and includes both decline (due to COVID-19 restrictions) and increases (after pandemic restrictions eased) in NTL at the (a) global, and (b) continental levels.}
    \label{fig:proppts}
\end{figure}

\begin{figure}
\centering
\resizebox{5.9in}{3.6in}{\includegraphics{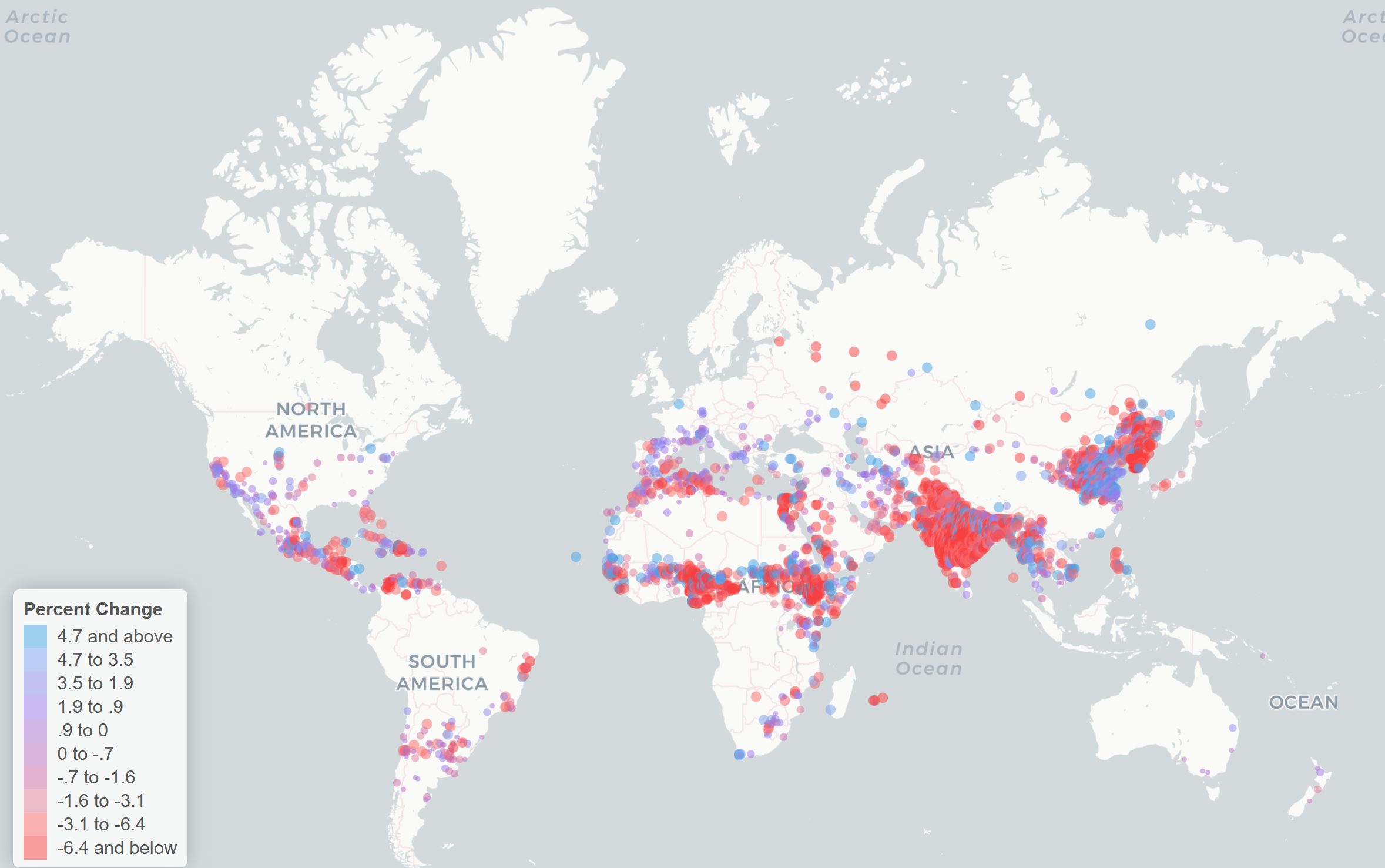}}
\vspace*{-.2mm}
\caption{Global map of NTL change severity on the date of highest NTL decline (April 6, 2020) after the onset of COVID-19, based on machine learning predictions. The markers show the percentage deviation between the machine learning predictions and incoming NTL data, indicating ongoing disruption due to response to policies and physical distancing (with negative change shown in red). Here we filtered the observations to only include urban areas with 50\% cloud-free minimum and 40\% gap-filled minimum quality to use high quality observations. }
\vspace*{-.5mm}
\label{fig:global_map}
\end{figure}

For the machine learning approach (2), our models~\cite{chakraborty2023adaptive-ts} learn the NTL dependence between consecutive windows from the pre-pandemic era for each urban area. Trained models are then applied to incoming data to monitor changes that are anomalous when compared with the natural variation in the data. This allows detection of any new changes of NTL even within the COVID-19 phase. Figure~\ref{fig:prA} shows the fraction of cities globally where anomalous change was detected based on deviation in observed NTL from machine learning predictions. Across all continents, there is a simultaneous rise in the proportion of urban areas globally where changes in NTL were observed starting from mid-March 2020, coinciding with the onset of COVID-19 induced restrictions. The proportion of change points continues to increase through 2022, because it is inclusive of both decreases in NTL from COVID-19 restrictions and subsequent increases in NTL due to resumption in normal urban activities after the pandemic restrictions eased. A similar pattern is visible at the continental scale (Figure~\ref{fig:prB}). Figure~\ref{fig:global_map} shows the global spatial representation of detected anomalous decreases in NTL on April 6, 2020, the date of the highest detected global average decrease in observed NTL compared to machine learning predictions. This
shows the detection of the global-scale disruption compared to preceding NTL data, shortly after the onset of COVID-19, with most urban areas showing a decrease in NTL of 6.4\% or more. The main advantage of the machine learning approach is that it can be useful for tracking short term responses to different policies enacted within the COVID-19 era and can highlight the detailed spatio-temporal variation of these responses globally.  

Figures~\ref{fig:sfig1}-\ref{fig:sfig6} show the daily distribution of the number of urban areas in each country where a decline in NTL was detected at each time step in 2020 using the machine learning models, for Africa, Asia, Europe, North America, Oceania, South America respectively, following the country codes from the ISO 3166 format. The per-country distribution is derived using kernel density estimates (with a Gaussian kernel, bandwidth = 1) of the number of urban areas in the country where a decline in NTL beyond the machine learning decision threshold was observed. In contrast to Figure~\ref{fig:prA} which shows fraction of urban areas where change was detected, Figure~\ref{fig:fig_dis} shows the per-country distribution of NTL decline (negative change) detected as a response to the pandemic — highlighting within continent heterogeneity in responses.
Distribution peaks indicate time periods when a large degree of urban areas in the country simultaneously show a trend of reduced human activity and NTL. As satellite observations are influenced by factors impacting retrievals such as cloud cover, for visualizing the results in Figures~\ref{fig:perchange}-~\ref{fig:fig_dis}, we applied data quality filters -- only selecting dates with a minimum gap-fill quality of 65\% and a minimum cloud free area of 50\% for each urban area (see Methods).  However, the full TRACE-NTL dataset consists of all observations, daily change measures associated with both approaches, and data quality measures, giving end users the flexibility to use the approach and quality thresholds that best fit their different applications. Given the lack of urban area-scale ground truth data on human activity variation during the pandemic, the trends observed in Figures~\ref{fig:perchange},~\ref{fig:rec_stl},~\ref{fig:proppts} serve as a confidence check with the global patterns corresponding with the timing of onset and easing of known pandemic restrictions. Additional validation of the machine learning approach is outlined in (Chakraborty
and Stokes, 2023)~\cite{chakraborty2023adaptive-ts}, and the decomposition approach is outlined in (Stokes and Roman, 2022)~\cite{stokes2022tracking}. 

\begin{figure}
\begin{subfigure}{1\textwidth}
  \centering
  \resizebox{7in}{9in}{\includegraphics[width=0.8\linewidth]{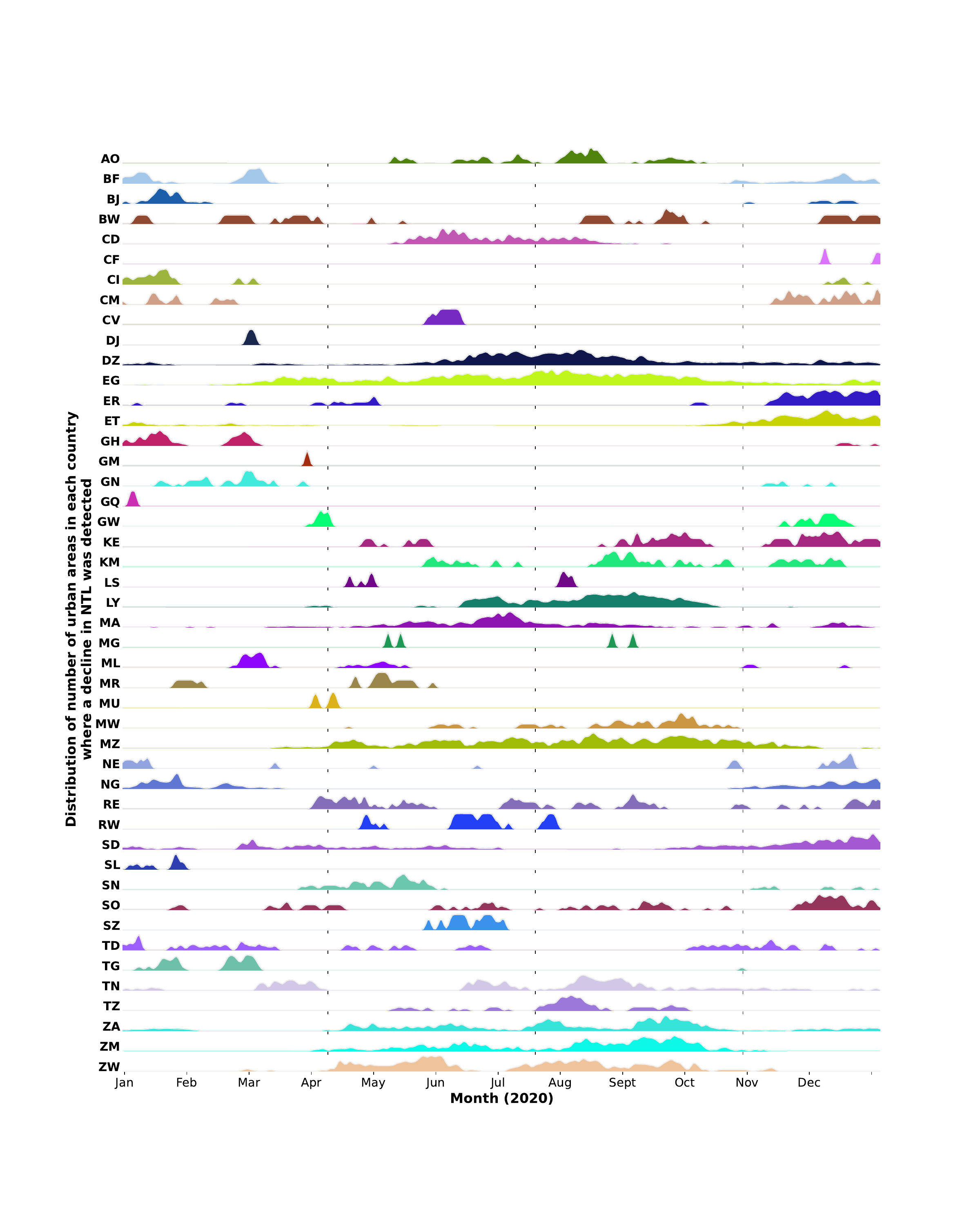}}
  \vspace{-6\baselineskip}
  \caption{Daily distribution of number of urban areas in each country in Africa where NTL drops were detected in 2020}
  \label{fig:sfig1}
\end{subfigure}%
\end{figure}
\begin{figure*}\ContinuedFloat
\begin{subfigure}{1\textwidth}
  \centering
  \resizebox{7in}{9in}{\includegraphics[width=0.8\linewidth]{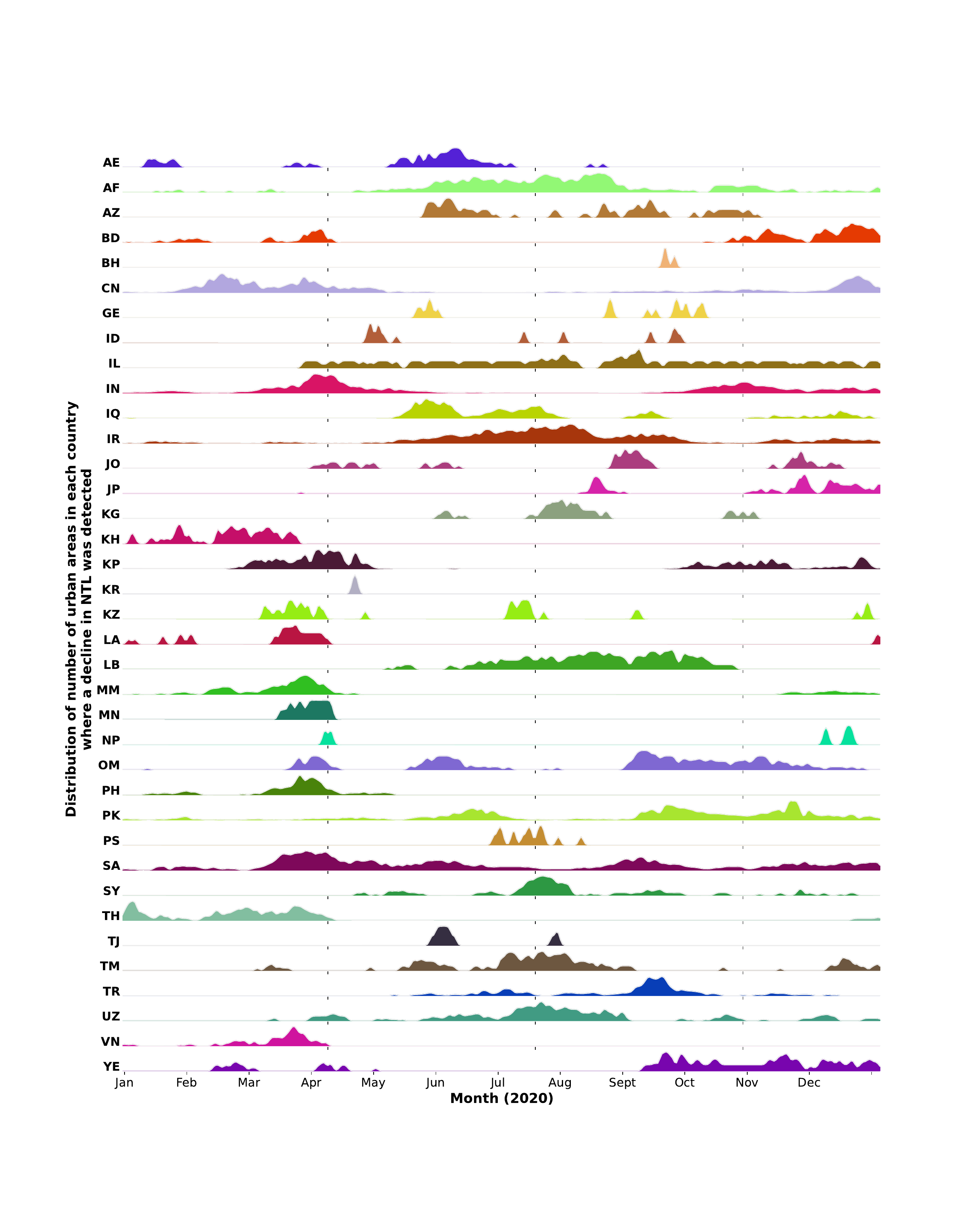}}
  \vspace{-6\baselineskip}
  \caption{Daily distribution of number of urban areas in each country in Asia where NTL drops were detected in 2020}
  \label{fig:sfig2}
\end{subfigure}
\end{figure*}
\begin{figure*}\ContinuedFloat
\begin{subfigure}{1\textwidth}
  \centering
 \resizebox{7in}{9in}{\includegraphics[width=1\linewidth]{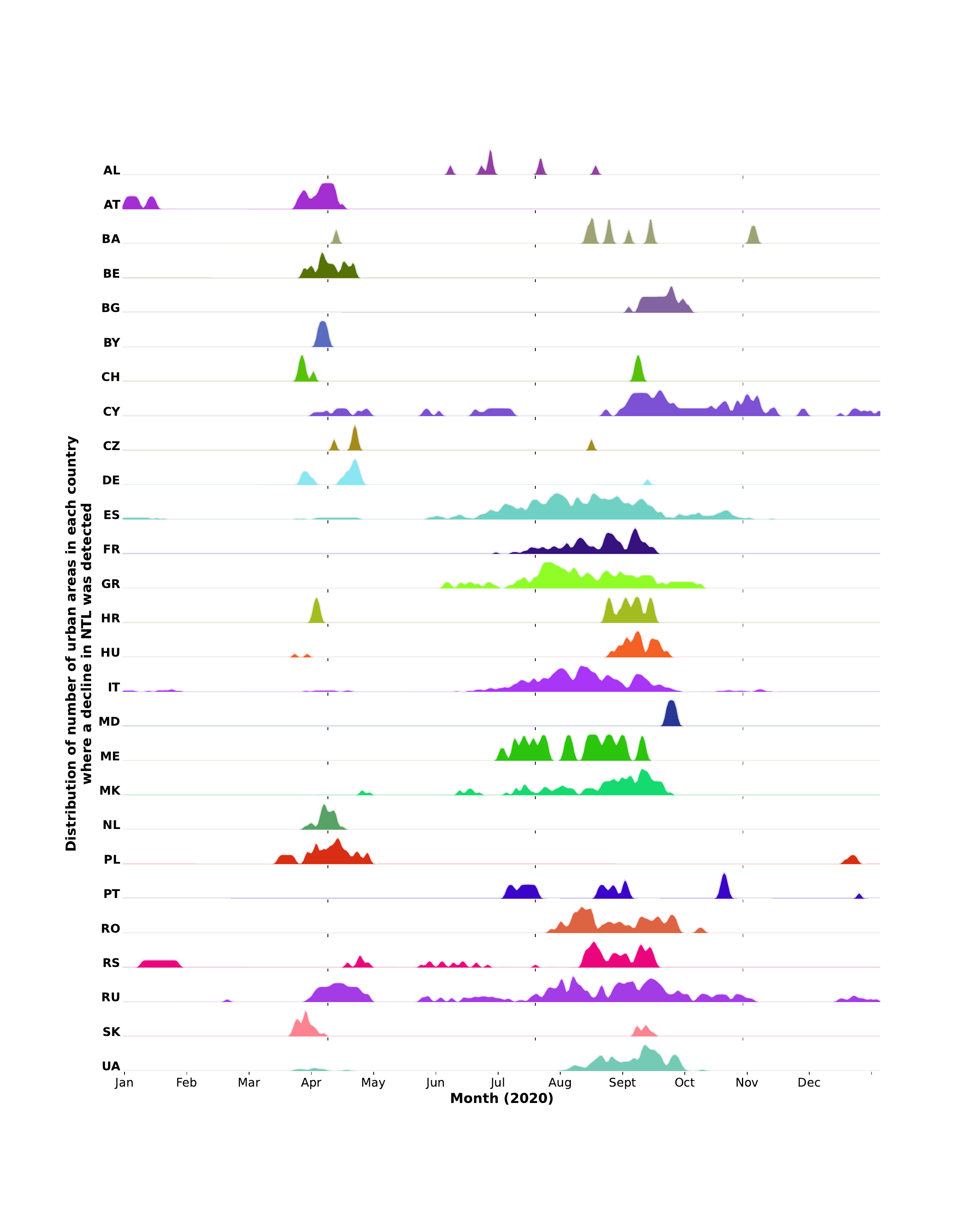}}
 \vspace{-6\baselineskip}
  \caption{Daily distribution of number of urban areas in each country in Europe where NTL drops were detected in 2020}
  \label{fig:sfig3}
\end{subfigure}
\end{figure*}
\begin{figure*}\ContinuedFloat
\begin{subfigure}{1\textwidth}
  \centering
  \resizebox{6.9in}{4.0in}{\includegraphics[width=0.8\linewidth]{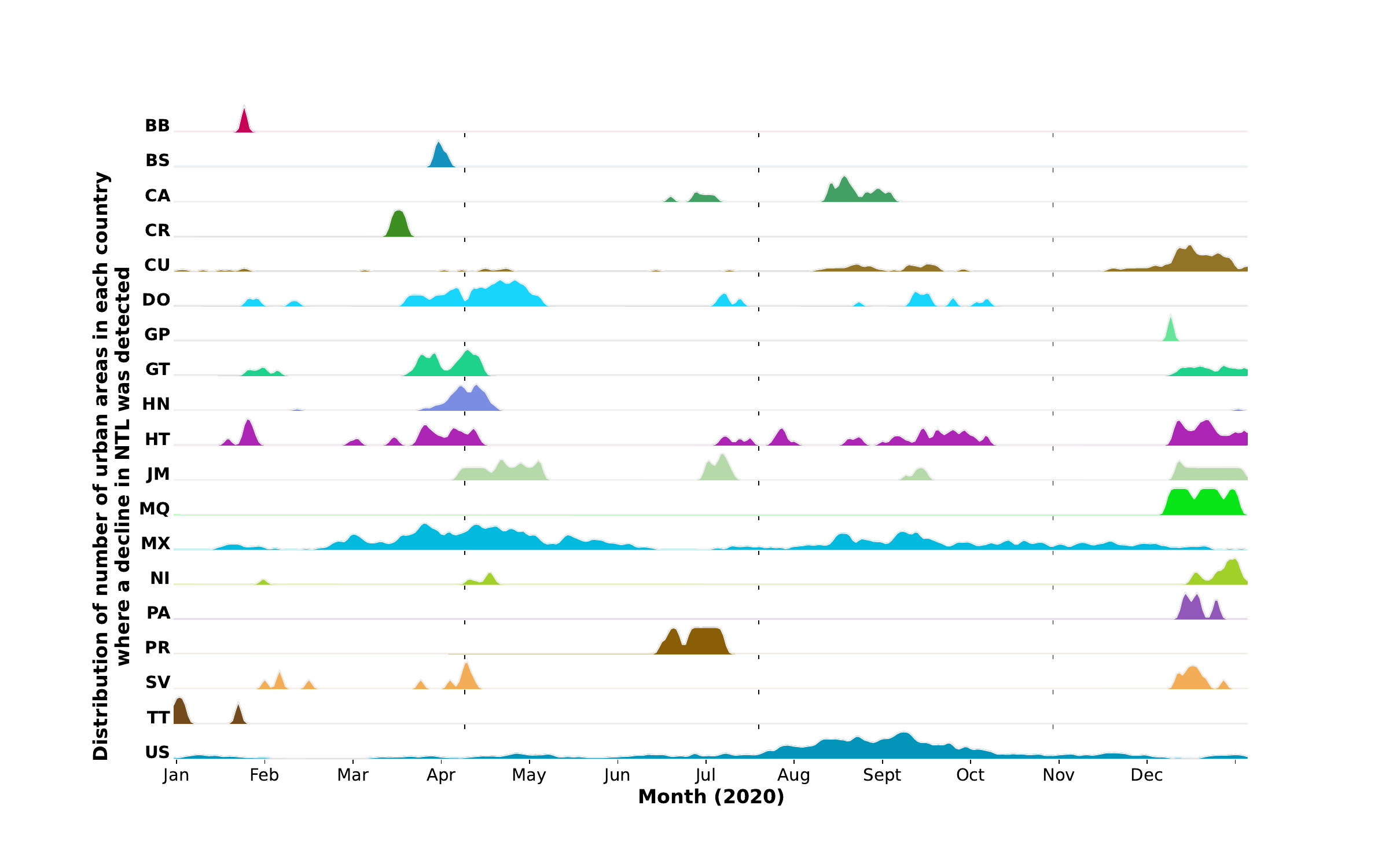}}
  \vspace{-3\baselineskip}
  \caption{Daily distribution of number of urban areas in each country in North America where NTL drops were detected in 2020}
  \label{fig:sfig4}
\end{subfigure}
\begin{subfigure}{1\textwidth}
  \centering
  \resizebox{6.7in}{1.8in}{\includegraphics[width=0.8\linewidth]{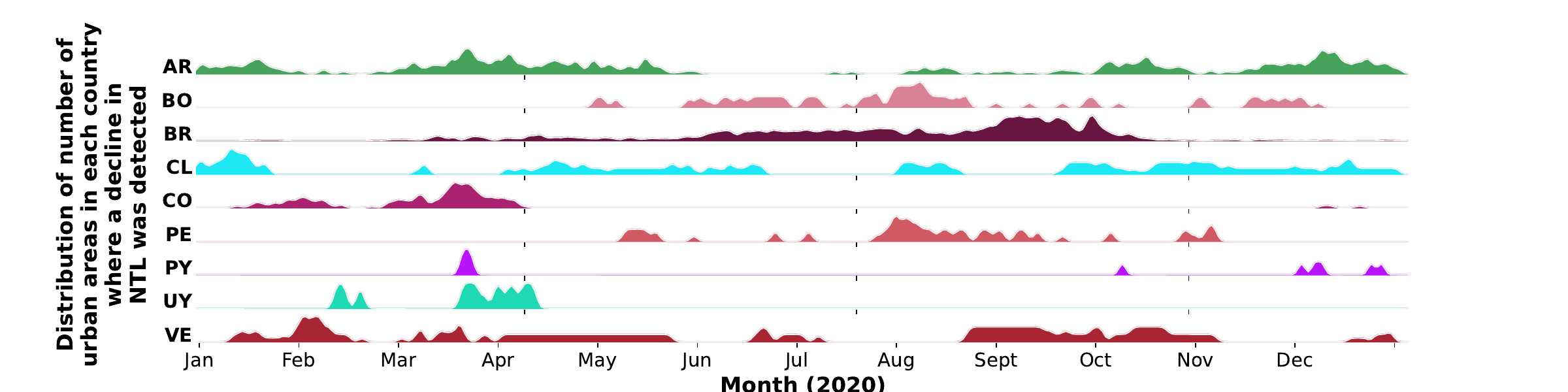}}
  \vspace{-0.1\baselineskip}
  \caption{Daily distribution of number of urban areas in each country in South America where NTL drops were detected in 2020}
  \label{fig:sfig5}
\end{subfigure}
\begin{subfigure}{1\textwidth}
  \centering
  \resizebox{6in}{1.6in}{\includegraphics[width=0.8\linewidth]{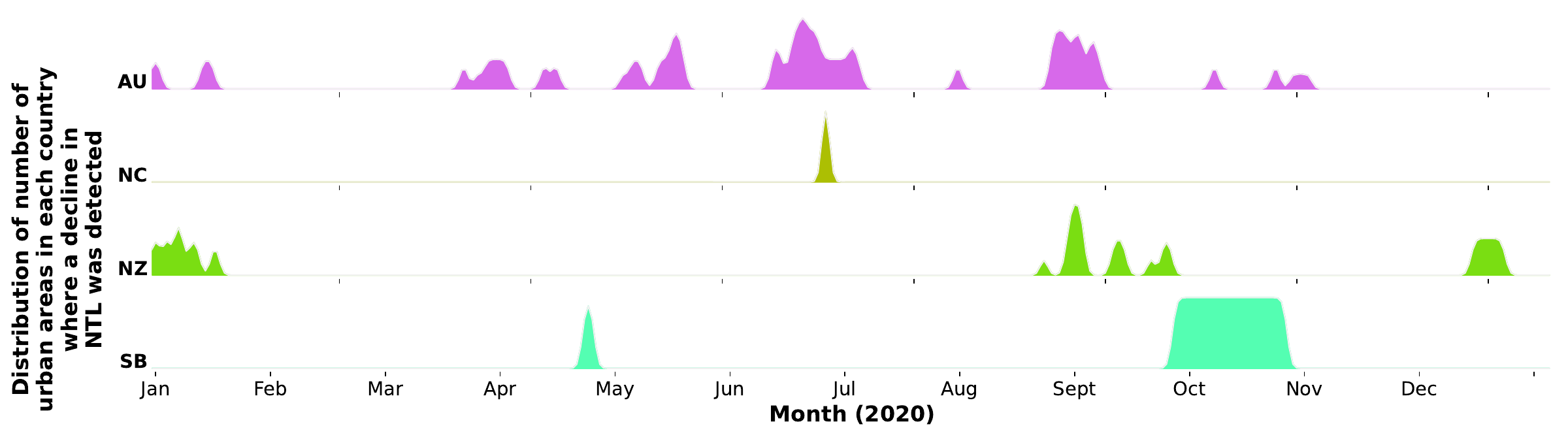}}
  \vspace{-0.5\baselineskip}
  \caption{Daily distribution of number of urban areas in each country in Oceania where NTL drops were detected in 2020}
  \label{fig:sfig6}
\end{subfigure}
\caption{Daily distribution of number of urban areas in each country where a decline (negative change) in NTL was detected in 2020 using machine learning, showing the changes in human activity during COVID-19 across each continent. The variation of the signal over time shows the spatial heterogeneity in NTL decline on an urban area level sub-nationally. The approach monitors deviations in NTL data stream and the temporal clusters show phases of ongoing NTL reduction. }
\label{fig:fig_dis}
\end{figure*}

\section*{Visualization Metrics Summary}
\label{ref:vis} 
An interactive visualization dashboard was created to display and aid in exploring the dataset. This tool can be accessed at \url{https://covid19-anthropause.nrp-nautilus.io/} and can be used to derive interactive global summaries of the results or to explore the full dataset in depth.  Users can view all urban areas on global map, zoom in to particular sub-regions, and explore the daily urban decreases in NTL detected by both approaches,
 The dashboard highlights dates when urban areas experienced anomalous dips in NTL, and also tracks how long NTL levels remain below predicted levels, and when recovery occurs. Users can specify "data frequency" ranging from 1-365 (in days) for viewing the dataset at different temporal intervals and "streak duration minimum" to define the minimum number of consecutive days of anomalous change to be included in the visualization results. Moreover, users can filter the results using the following data quality indicators:
\begin{enumerate}
    \item Cloud free urban area minimum (\%): minimum percent proportion of the urban area that is cloud free on a particular day. Cloud free urban area and ranges from 0 - 100, with 0 being completely cloud covered at that time-step.
    \item Gap fill quality minimum (\%): minimum percent proportion of days in a 30 day time period centered on the day of choice with good quality observations (VIIRS Black Marble QA flag 0 or 1). Gap filled quality ranges from 0 - 100, with 0 being all observations in 30 day period are gap filled or interpolated.
\end{enumerate}

The dashboard allows users to track the timing and severity of anomalous urban NTL decreases through the “disruption tab (using the machine learning approach), and also the timing and extent of recovery through the “recovery” tab (using the decomposition approach). These maps of NTL changes are displayed at user-defined
temporal intervals (in days), gap filled minimum \%, and cloud cover minimum \%. The map can be viewed with urban areas indicated as spatial markers or as a chloropleth, with the color scale of the map indicating the intensity of the change in the urban area on a given date through percent change. Users can also interact further with the visualization by clicking the urban area locations (markers or chloropleth) to display the country, urban area name and code, and the quantitative percentage change value on the selected date. In addition, the dashboard  displays individual urban area time-series data through the “plots” function, showing NTL observations, the predicted NTL from machine learning ensembles, and the detected change points. Users can select the country, cities in the dataset, and the corresponding site ID, and submit these preferences for display. A "cloud free urban area minimum" and "gap filled quality minimum" selection of 0 creates the time-series display from all time-steps used in deriving the predictions.

\section*{Methods}
Our study is centered around deriving spatio-temporal (daily, global at the urban area level) metrics from NTL data and consists of initial acquisition and refinement of NTL imagery, followed by analysis. For analyses, we derive (i) a set of disruption metrics that describe how NTL deviates from its expected behavior, (ii) recovery metrics that describe any observed increases in NTL indicating recovery from pandemic disruptions, and (iii) confidence and data quality metrics that show decision confidence based on agreement between multiple machine learning models as well as the stability of the baseline time-series to show decision uncertainty metrics and impact of cloud cover, gap filling on input data quality. The proposed workflow is described in Figure~\ref{fig:workflow}.

\subsection{Data acquisition and refinement}
\begin{figure}
\centering
\resizebox{5.75in}{4.75in}{\includegraphics{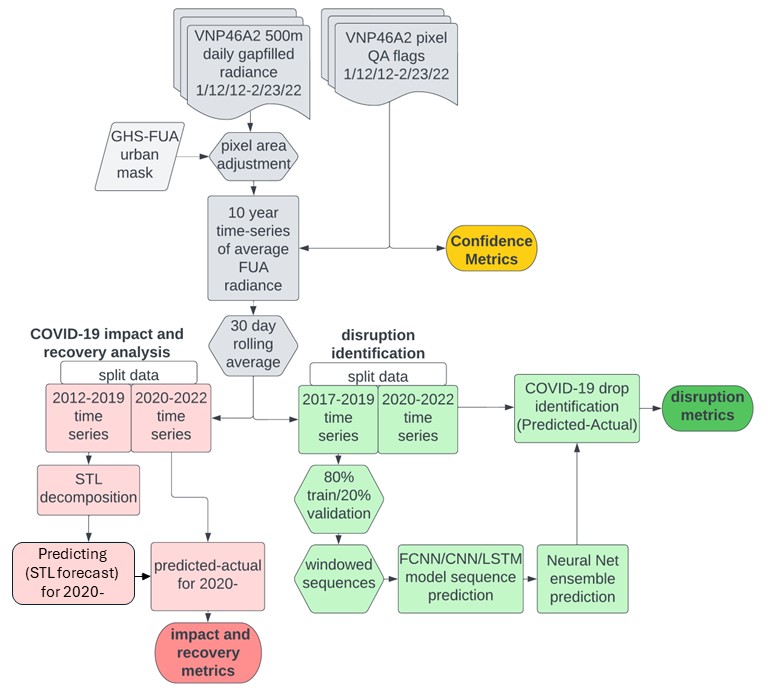}}
\vspace*{-.2mm}
\caption{Workflow of the proposed methodology to derive COVID-19 disruption and recovery metrics from NTL time series. Pre-processing steps to derive a 1-D NTL time series in grey; pre-COVID comparison analysis in pink; disruption identification analysis in green. }
\vspace*{-.5mm}
\label{fig:workflow}
\end{figure}

We use the VNP46A2 dataset from NASA’s Black Marble Product Suite that records daily BRDF-corrected nighttime lights from the Day/Night Band onboard the VIIRS instrument~\cite{roman2018nasa} at 500m to construct urban NTL time-series over a period of 10 years for each urban area for the period of January 12th 2012 until February 23 2022, employing the entire data archive at the time we began our analysis (Figure \ref{fig:workflow}).
The product was accessed via NASA's Level-1 and Atmosphere Archive and Distribution Active Archive Center (LAADS-DAAC) System Web Interface.  This product provides daily atmospheric-, terrain-, vegetation-, snow-, lunar-, and stray light-corrected radiance at night, minimizing the influence of extraneous artifacts in the final NTL values.  To keep the sampled pixels in each urban area constant from day to day, we chose the gap-filled BRDF-corrected layer of the VNP46A2 dataset, which uses temporal gap-filling to replace cloud-covered pixels with historically valid observations to reduce persistent data gaps. By using the gap-filled layer, our time series is biased toward stability, and observed decreases in light are a conservative indication of actual anomalies on the ground.

All grid tiles for the product were downloaded, constituting global coverage.  We masked each tile to only include urban areas, using the Joint Research Commission’s Global Human Settlements Functional Urban Areas dataset (GHS-FUA) \cite{schiavina2019ghsl}.  GHS-FUA is the most complete global database to date of the locations, names, and spatial extent of thousands of human settlements and their commuting zones \cite{moreno2020metropolitan}. For the pixels in each urban polygon, we first remove pixels with gap-filled NTL matching the VNP46A2 fill value and for the remaining pixels, we sum the radiance measurements weighted by each pixel's area and divide by the total estimated area of all utilized pixels to give an area-weighted radiance intensity measurement in units of nW cm$^{-2}$ sr$^{-1}$. The estimated area of each pixel is based on a WGS84 datum.  We create a time-series from the average daily radiance for each functional urban area, and utilize a 30 day rolling average, where the radiance at a given time-step is the average of the 15 observations leading and trailing it. This creates daily 1-D time-series of each urban area and is used in our study to identify the dates of COVID-19 disruptions for assessing how and when it impacted urban NTL, and to record time periods when NTL recovers to normal levels. We also extract the ``Mandatory Quality Flag'' from the VNP46A2 dataset to compute daily quality assessment measures on minimum cloud free area and minimum gap-filled area. 
\subsection{Disruption date and magnitude identification}
\subsubsection{Methods: Unsupervised Deep Learning-Based Forecasting and Time-Series Analysis}
We adapt a deep learning-based forecasting approach \cite{chakraborty2023adaptive-ts} (approach 2 described before) to detect disruptions in urban activity captured by NTL, by monitoring the deviation of the observed NTL from model forecasts. This involves learning a function $f$ that characterizes the temporal dependence between consecutive sequences of NTL observations from pre-COVID baseline years. For an input sequence, $X_i=[x_{t-w_i},\dots, x_{t-1}]$, the function learns to predict the next temporal sequence $X_o=[x_{t},\dots, x_{t+w_o}]$ at time-step $t$ as $X_o=f(X_i)$, where $w_i$ and $w_o$ are the input and output temporal windows and $x_t$ denotes the observed NTL at $t$. The models learn the stable behavior from the training phase characterizing the expected city-specific dependence between NTL levels in input and output windows. The trained model is applied on unseen test data to predict the output NTL window. In the absence of change, this function predicts the output NTL sequence given the preceding input window, with a low error. During change, the observed NTL deviates from the expected behavior predicted by the model, resulting in a high prediction error that is monitored to detect change. We used 60 days and 30 days for $w_i$ and $w_o$ respectively, which were determined experimentally over a subset of randomly selected 200 urban areas to best capture gradual changes associated with COVID-19 disruptions and kept constant for the study. Given the length of the output sequences, each $x_t$ is forecasted 30 times, and we take the median of these forecasts as the resultant prediction, denoted as $\hat{x}_t$.  

The approach is unsupervised and learns the dependence from the natural variations in NTL in a city, allowing the use of a large volume of unlabeled data for training. Training sequences consist of NTL data from 2017 to 2019 to learn the temporal dependence from the years immediately before COVID-19, using a 80\%, 20\% split for training and validation as shown in Figure~\ref{fig:workflow}. Although non-COVID-19 related disturbances may exist during the training phase, these disturbances constitute only a small fraction of the data and hence does not affect training, except in the rare cases of long term loss of NTL,.e.g. from disasters or geo-political conflicts. 

We use multiple neural network architectures suited for sequential data~\cite{goodfellow2016deep} to learn the forecasting functions: 1-D convolutional neural network (1-D CNN), fully connected neural networks (FCNN), and recurrent architectures based on the long short term memory (LSTM) network~\cite{hochreiter1997long}. Together, these represent the widely used models for time-series analysis and encompass the fundamental architectures for studying sequential data~\cite{ismail2019deep}. The trained models are applied to test phase data (2020-2022) to derive predictions for the respective urban area and we create an ensemble prediction from these three models for a time step as:
\begin{equation}
    \hat{x}_{t,\text{ens}}=w_{\text{cnn}}\hat{x}_{t,\text{cnn}}+w_{\text{fcnn}}\hat{x}_{t,\text{fcnn}}+ w_{\text{lstm}}\hat{x}_{t,\text{lstm}},
    \label{eqn:ens}
\end{equation}

where $w_{\text{cnn}}, w_{\text{fcnn}}, w_{\text{lstm}}$ are the ensemble weights 0.2, 0.3, and 0.5 respectively. These weights were determined experimentally over the sample of 200 urban areas by considering the model behavior and robustness and kept constant for the study. FCNN and LSTM networks outperform 1-D CNN at adapting to NTL variations, with best performance observed with LSTM. 

To identify disruptions, the ensemble prediction at time step $t$ from equation~\eqref{eqn:ens} is compared with observed NTL $x_{t}$ to compute the squared residual at $t$ as
\begin{equation}
    \epsilon_{t}=(x_{t}-\hat{x}_{t,\text{ens}})^2.
    \label{eqn:err}
\end{equation}

High error $\epsilon_{t}$ indicates that $f$ fails to model the NTL variation based on past knowledge, and the time-step is associated with disruptions in urban activity. The change detection threshold $\tau$ is set so that it detects the top $25 \%$ error time steps for each urban area. All time-steps with $\epsilon_{t}>\tau$ are marked as change points. 
We used TensorFlow/Keras and Python for training and utilized graphical processing units (NVIDIA-GeForce-RTX-3090) based distributed processing using the Nautilus HyperCluster on the National Research Platform~\cite{nrp} for accelerating training. We used the same architecture and parameters as described in~\cite{chakraborty2023adaptive-ts} for all urban areas, showing the robustness of the configurations to adapt to city-specific trajectories.

\subsubsection{Time-series metrics to quantify disruptions in urban activity caused by COVID-19}
\label{sec:dis}


Using the ensemble predictions from equation~\ref{eqn:ens} and detected change points by thresholding equation~\ref{eqn:err}, we derive the following metrics to characterize change at every time-step throughout the study duration from 2017-2022. 
\begin{enumerate}
    \item magnitude of change showing severity of deviations
\begin{equation}
    m_{t}=|x_{t}-\hat{x}_{t,\text{ens}}|.
    \label{eqn:met_mag}
\end{equation}
    \item direction of change from baseline and preceding window
\begin{equation}
    d_{t}=sgn(x_{t}-\hat{x}_{t,\text{ens}}),
    \label{eqn:met_dir}
\end{equation}
where $sgn(.)$ indicates the sign of the residual. Thus, $d_{t}<0$ implies a decrease in NTL at $t$ (negative change, for e.g. due to reduced urban activities during COVID-19 as a result of interventions such as physical distancing) and $d_{t}>0$ indicating an increase in NTL (positive change, for e.g. due to recovery as physical distancing restrictions are reduced). This is computed only for detected change time-steps ($\epsilon_{t}>\tau$).  
    \item change decision: a binary metric (change:1, no-change:0) indicating whether or not the time-step $t$ is a change point determined by noting high prediction error $\epsilon_{t}>\tau$.
    \item detection confidence: average agreement among the predictors while computing the ensemble result (1 indicating complete agreement).
\end{enumerate}

We then consider successive change time-steps detected by the ensemble as change segments, each having a start $t_{\text{start}}$, and end $t_{\text{end}}$, showing the change duration and compute the following segment specific metrics:
\begin{enumerate}
    \item overall magnitude or severity of change across the segment compared to $x_{t_{\text{start}}-1}$, the NTL observation preceding the segment.
\begin{equation}
    \mathit{M}_{\text{seg}}=\sum_{k=t_{\text{start}}}^{t_{\text{end}}} |x_{t_{\text{start}}-1}-x_k|.
    \label{eqn:met_smag}
\end{equation}
    \item average magnitude of change across the segment, showing the severity of change given the segment duration, or the average change severity per time-step of the change segment
\begin{equation}
    \mathit{A}_{\text{seg}}=\frac{\sum_{k=t_{\text{start}}}^{t_{\text{end}}} |x_{t_{\text{start}}-1}-x_k|}{t_{\text{end}}-t_{\text{start}}+1}
    \label{eqn:met_amag}
\end{equation}
    \item segment inflection point $t_i$ showing time-step of global maximum of prediction error magnitude in the segment and marks the point where the model adjusts to the change with incoming data, causing the prediction error to reduce. This is determined by computing the residual at every time step $t$ in the change segment as $r_t= |x_t-\hat{x}_{t,\text{ens}}|$ and selecting $t_i$ with highest $r_t$ as
    \begin{equation}
    t_i=\arg \max_{r}([r_{t_{\text{start}}}, \dots, r_{t_{\text{end}}}]).
    \label{eqn:met_inf}
\end{equation}
    \item change start rate quantifying the rate of change in NTL to its maxima over the change segment as 
    \begin{equation}
    \mathit{R}_s=\frac{\hat{x}_{t_i}-\hat{x}_{t_{\text{start}}-1}}{t_i-t_{\text{start}}}
    \label{eqn:met_rinc}
\end{equation}
    In change segments over which NTL reduced in magnitude during COVID-19 restrictions, this measure shows how rapidly NTL declined as a response to policies.
    \item change end rate quantifying the rate of recovery of NTL from the change maxima in the segment to pre-change NTL levels for the segment. 
    In change segments where NTL decreased in magnitude due to physical distancing measures, this serves as measure of how rapidly NTL resumed to pre-change levels due to potential reduction such restrictions and is derived using
     \begin{equation}
    \mathit{R}_e=\frac{\hat{x}_{t_{\text{end}}-1}-\hat{x}_{t_i}}{t_{\text{end}}-t_i}
    \label{eqn:met_dec}
\end{equation}
\end{enumerate}
Together these metrics capture the disruption caused by COVID-19 across cities as seen from drops in NTL. 
\subsection{Decomposition approach to quantify recovery in urban activity levels}
\label{sec:rec}
Since the neural network ensembles use sliding windows to derive predictions, the ensemble predictions adjust to evolving trends  in the time-series from preceding windows. Therefore, with the machine learning approach, change in NTL is measured against a constantly evolving baseline. To derive predictions compared to a single pre-COVID baseline, we decompose the NTL 30 day rolling average time-series data from 2012-2019 using the Seasonal and Trend decomposition using Loess (STL) into trend, season and residual components~\cite{cleveland1990stl}, following the approach outlined in previous studies~\cite{stokes2022tracking}. NTL has been shown to have a strong seasonal signal in some parts of the world associated with holidays and seasonal migrations~\cite{roman2015holidays}, so to include those cultural processes in the NTL baseline, we assume the seasonal component repeats each year. We then forward forecast the trend and seasonal components and sum them to formulate a prediction for 2020-2022 for each urban area as $\hat{x}_{t,\text{stl}}$. 
We measure recovery from COVID-19 throughout 2020-2022 by comparing the observed NTL with the STL-based forecasts during this phase. Recovery is marked by time-steps with $x_t-\hat{x}_{t,\text{stl}} >0$, where the observed NTL exceeds the STL forecast and we compute this difference per time-step (between observed NTL and forecast)  as
\begin{equation}
    \mathit{D}_t=x_t-\hat{x}_{t,\text{stl}}.
    \label{eqn:met_stl_rec}
\end{equation}
We denote recovery state per time-steps with $\mathit{D}_t>0$ as marked as 1 (recovery) and 0 otherwise (no recovery).

For both ensemble and STL predictions, we compute percentage change $p$ as the deviation of observation from the predicted NTL using
\begin{equation}
    \mathit{p}_t=\bigg(\frac{\hat{x}_t-x_t}{\hat{x}_t}\bigg)\times 100,
    \label{eqn:perc}
\end{equation}
where $\hat{x}_t$ is the ensemble or STL prediction at the time step and the direction of change noted as $sgn(-p_t)$, with declines marked as negative.
\subsection{Uncertainty and quality assessment}
\label{ref:unc}
As neural network change or anomaly detectors model the expected baseline variation to then detect deviations from it, the quality or stability of the training data that is used to learn baselines, plays an important role in determining the deviations that will stand out to the model as anomalies. To quantify the variability in the training data, we derive uncertainty metrics based on the training data for each urban area. In regions with high NTL variability, only very large disruptions can be confidently detected by the methods. Data users can rely on uncertainty metrics to interpret how likely it is to detect disruptions in each urban area, given the natural variability of the NTL time-series. The following metrics are computed for each urban NTL time-series over all time-steps during the training period from 2017-2019 with $k$ observations. 
\begin{enumerate}
    \item coefficient of variation (cv) that measures the dispersion of the training set NTL time-series about its mean over each urban area as
    \begin{equation}
    \text{cv}=\frac{\sigma(x_{k})}{\mu(x_{k})}.
    \label{eqn:pv}
\end{equation}
    \item proportional variability (pv)~\cite{heath2013quantifying} based on pairwise ($i, j$) comparison of ratio of all $C=\frac{k(k-1)}{2}$ NTL time-steps in the training set and has been shown to eliminate the bias in the cv measure by being robust to outliers and not requiring any assumptions on underlying statistical distributions. This is computed using
    \begin{equation}
    \text{pv}=\frac{1}{C}\sum \frac{|x_{i}-x_{j}|}{max(x_{i},x_{j})}.
    \label{eqn:pv}
\end{equation}
    \item consecutive disparity index (cdi) ~\cite{fernandez2018consecutive} that also accounts for the chronological ordering of observations and quantifies the average rate of change in consecutive values in the NTL training set as a measure of variability. Like pv, cdi is also not dependent on the mean of the training set, is robust to outliers while taking into account the chronological order of the data. This is computed using
    \begin{equation}
    \text{cdi}=\frac{1}{(k-1)}\sum_{i=k}^{k-1} \bigg |ln \frac{x_{i+1}}{x_{i}} \bigg |.
    \label{eqn:pv}
\end{equation}
\end{enumerate}
These metrics capture the stability of the NTL from training phase and are useful for interpreting ensemble decision reliability.
In addition, we also derive three quality control metrics for each time-step that can be used to filter the disruption date results from poor quality retrievals, or short-lived disturbances based on the quality flags of the NTL observations in the Black Marble product: cloud free \%, gap fill quality \%, and streak duration as discussed with our visualization tool description. These indicators are used to assess the data retrieval quality per time-step. 
Cloud free \% indicates the percentage of the urban area that is cloud free at each date. This is determined from Black Marble data quality layers. 
Gap fill quality \% indicates the minimum proportion of days in a 30 day time period centered on the day of choice with good quality observations and is determined from Black Marble data quality layers. 
Streak indicates the length of the disruption--i.e. the number of disruption dates in a row. Filtering by streak length allows the data user to weed out NTL drops that may be caused by short term events, like intermittent power outages. 
These data quality metrics are derived for each urban area and shows the stability of the training data per time-step and are useful for interpreting both the ensemble and STL decisions.
These are provided with the dataset for assessing the reliability of the input data and consequently the model decision. Moreover, the change detection confidence derived from the ensemble of neural networks shows the agreement to also denote trust in model decisions. 

\subsection*{Code availability}

To derive our analyses for the project, we used TensorFlow/ Keras (version 2.6) in Python (version 3.9). The project code to derive the disruption, recovery metrics is hosted on project GitHub from \url{https://github.com/srijac/covid-19_Nightlights}. This dataset~\cite{trace-ntl} is available on Zenodo at \url{https://zenodo.org/records/11206064}.

\section*{Data Records}


The TRACE-NTL dataset consists of input NTL data to the models, machine learning and STL predictions (daily, global) and the derived metrics describing disruption caused by COVID-19 from the ensemble-based change statistics of each time-step, change segment information, comparison with normal pre-COVID years using STL to note recovery at each time step, decision confidence, and uncertainty and data quality measures. These measures describing the impact of COVID-19 on NTL levels are stored in .csv and .txt format for each urban area. This format allows it to be easily used in statistical software packages for data compilation or analysis. The dataset consists of three main directories $\texttt{data}$, $\texttt{metrics}$ and $\texttt{ancillary}$. The $\texttt{data}$ directory consists of NTL input to the ensembles and the output from machine learning analysis per time-step for each urban area. The directory $\texttt{metrics}$ contains two further hierarchies on machine learning-derived $\texttt{disruption}$ and STL-derived $\texttt{recovery}$ quantifications. The $\texttt{ancillary}$ directory has two files: (a) GHS\_FUA\_ID\_Country.csv that indexes all 9031 GHS urban areas to its ID ($\texttt{fua\_ID}$), its Black Marble tile ($\texttt{fua\_tile}$), name and country, and (b) country\_ISOcodes2letter.csv describing the two letter country codes used in the Figure~\ref{fig:fig_dis}.
For each urban area the TRACE-NTL dataset consists of five sets of metrics describing COVID-19 induced NTL changes in files under the $\texttt{metrics}$ directory. Further description on this is provided below:
\begin{enumerate}
    \item \textbf{disruption indicator at each time-step}: These files describe change statistics of each time-step showing date, observed NTL, ensemble predicted NTL, magnitude of NTL deviation from ensemble prediction, the direction of deviation, binary change decision, percentage change, decision confidence derived from the neural network ensembles from 2017-2022 listing the metrics described in Section~\ref{sec:dis} in .csv format. For each urban area, this file is found in $\texttt{metrics/disruption/daily\_change/}$. These are derived from the files in the $\texttt{data}$ directory and follow the convention $fua\_<fua\_ID>\_<fua\_tile>\_time\_step\_change\_metrics.csv$.
    \item \textbf{disruption statistics of change segments}: For consecutive anomalous detections, the dataset consists of change segment information showing start date, end date, inflection point, rate of change increase, decrease, total change magnitude and average magnitude of change over the segment from neural network ensembles from 2017-2022 describing the metrics from Section~\ref{sec:dis} in a .csv format. For each urban area, this file is found in $\texttt{metrics/disruption/change\_segment}$. These are derived from the files in the $\texttt{data}$ directory and follow the convention $fua\_<fua\_ID>\_<fua\_tile>\_segment\_change\_metrics.csv$.
    \item \textbf{recovery indicator at each time-step}: Recovery metrics based on the difference between STL forecast and observed NTL for 2020-2022 and recovery state are shared as .csv files listing the metrics described in Section~\ref{sec:rec}. Each file consists of dates from 2020-2022, the rolling average NTL, STL forecast, difference (dataset shows forecast-observed), state showing whether recovery condition from equation~\ref{eqn:met_stl_rec} is met at the time-step. For each urban area, this file is found in $\texttt{metrics/recovery}$, follow the convention $fua\_<fua\_ID>\_<fua\_tile>\_recovery\_metrics.csv$ and the analysis is derived from data and forecasts shared in these files.
    \item \textbf{uncertainty metrics (cv, pv, cdi)}: These metrics are shared as a .txt file described in Section~\ref{ref:unc} capturing the variability of each urban area's NTL time-series over the training period from 2017-2019. For each urban area, this lists the three metrics computed from the training duration and the files are found in $\texttt{metrics/disruption/city\_uncertainty}$ and follow the convention $fua\_<fua\_ID>\_<fua\_tile>\_U\_Output.txt$. 
    \item \textbf{data quality indicators}: minimum cloud free urban area and minimum gap-fill quality (as percentages) at each time-step from 2017-2022 are shared in a .csv format capturing the data quality indicators described in Section~\ref{ref:unc}. For each urban area, this file is found in $\texttt{metrics/disruption/daily\_qa\_flags}$. These are derived from Black Marble mandatory quality flags and follow the convention $fua\_<fua\_ID>\_<fua\_tile>\_quality\_metrics.csv$.
\end{enumerate}

\section*{Technical Validation}



Given the spatial (global) and temporal (daily) scale of the dataset over several years, we utilized a combination of visual and quantitative approaches to assess the dataset's usability in informing human activity change caused by COVID-19 for downstream studies. Visually, we examined the models' behavior and metrics across a random selection of approximately 500 urban areas, which is about 5.53\% of the global cities in the dataset. The models perform as expected by tracking gradual trends and detecting relatively abrupt deviations if they are anomalous compared to the pre-COVID baseline and the metrics derived are thus consistent. Quantitatively, we have also used an ensemble of neural networks and noted the agreement between models to denote decision confidence. We observe a high agreement between models which is an indicator of reliable decision and metrics at those time-steps. These predictions also depend on the quality of training data and incoming test data and we denote the reliability of these factors through uncertainty metrics and data quality indicators. Data quality measures of minimum cloud free percentage, gap fill quality, duration of change are thus important quality control measures for the dataset. In addition, the global and continental scale aggregates of percentage change shows change in NTL coinciding with the onset of COVID-19 (Figure~\ref{fig:perchange}), the fraction of urban areas undergoing change increases with the onset of COVID-19 and continues over time as normal activities resume resulting in increased NTL (Figure~\ref{fig:proppts}), and a global recovery trend from the fraction of global urban areas with NTL exceeding the forecast (Figure~\ref{fig:rec_stl}), showing resumption of activity levels with the ease of restrictions. These show that the change measures in the dataset agree with the event and can serve as a indicator of human activity change (through NTL) to further understand the local and temporal variations of the COVID-19 anthropause as captured in Figure~\ref{fig:fig_dis}. 
\section*{Usage Notes}



The dataset is shared as multiple .csv files and a .txt file for compatibility with standard data processing and analysis software and packages to facilitate a wide-range of downstream studies by researchers and stakeholders for examining the disruption caused by the pandemic on human activity as captured by NTL and the impact of this disruption, in turn, on various environmental and economic variables. For each urban area, the .csv files consist of disruption metrics showing the date, NTL radiance, predictions and per time-step metrics described in Section~\ref{sec:dis} showing magnitude of prediction error, binary indicator of change point, change severity and direction. Disruption metrics for the change segments including start and end dates, rates, and inflection points described in Section~\ref{sec:dis} are also provided. These disruptions are derived with respect to incoming data and show short-term responses to policies and continually adjust to ongoing changes. Recovery metrics of each urban area (.csv files) from Section~\ref{sec:rec} show if the NTL observation is higher than forecasts derived from pre-pandemic years. The forecasts are derived with a static representation of normal years and show deviations from it, but does not account for ongoing changes 2020 onwards.  For interpreting both sets of metrics and its reliability, use of uncertainty measures and daily data quality indicators are important and highlight the NTL variability in that urban area. For areas with high variance in baseline, only severe anomalous changes can be confidently detected. Urban areas impacted frequently by poor quality retrievals, may not have a stable baseline and only large deviations in such areas can be detected with high confidence. Per time-step gap-filled and cloud cover percent metrics, baseline stability in Section~\ref{ref:unc} provide these metrics and help in correctly assessing how severe COVID-19 related changes are given the long-term behavior of the area. While using the dataset, noting these metrics are important to estimate how trustworthy the decisions are. 

An example usage of these disruption and recovery metrics would be using the NTL time-series of an urban area of interest, the detected disruption dates, severity and examining how well it follows physical distancing policies and other social measures introduced in the area. Correlation of reduced human activity and recovery of activity with various environmental and public health variables can also be examined due to the high temporal resolution of the dataset. The disruption metrics for the change segments can illustrate how human activity in an urban area is impacted by successive updates to physical distancing mandates and its effect on a given city. 
Although, NTL does not directly relate to human activity, it relates to presence of urban infrastructure and captures certain aspects of changes in human activity such as business closures, changes in use of public space, changes in traffic, all of which have been impacted by COVID-19 physical distancing policies and our dataset captures these variations through NTL. 
Overall, we expect our dataset to provide high temporal resolution urban area scale metrics to quantify these changes from satellite observations of nighttime lights as a response to COVID-19 and fill an important gap that records locally relevant global scale change information daily to support further analysis linking them to changes in policy, climate and economic variables.

\section*{Acknowledgements}
Funding for this project was provided by NASA's Rapid Response and Novel Research in Earth Science (RRNES) grant 80NSSC20K1083. Computational resources on the National Research Platform was obtained through the Research Institute for Advanced Computer Science, USRA.

\section*{Author contributions}
S.C. developed the time-series analysis and metrics, lead the analysis execution, contributed summaries of the results, developed the dataset, and contributed to the writing.
E.C.S. obtained funding for the project, conceived the research question and design, contributed analysis and summaries of the results, contributed to dataset development, and contributed to writing the manuscript.
O.A. developed analysis plots.
E.C.S and O.A. developed the global COVID-19 visualization
All authors reviewed and revised the manuscript. 

\section*{Competing interests}
The authors declare no competing interests.

\bibliographystyle{unsrt}  
\bibliography{references}

\end{document}